\documentclass[%
 reprint,
 superscriptaddress,
nofootinbib,
 amsmath,amssymb,
 aps,
 prd,
floatfix,
twocolumn,
notitlepage
]{revtex4-1}
\usepackage{bbm}
\usepackage{lipsum}
\usepackage{graphicx}
\usepackage{dcolumn}
\usepackage{subfigure}
\usepackage{bm}
\usepackage{breakurl}
\usepackage{enumitem}
\usepackage{xcolor}
\usepackage[normalem]{ulem}
\usepackage{multirow}
\usepackage[multidot]{grffile}

\definecolor{myblue}{rgb}{0.153,0.322,0.706}
\usepackage[colorlinks,linkcolor=myblue,urlcolor=myblue,citecolor=myblue]{hyperref}
\usepackage{geometry}
\newcommand{\be}{\begin{equation}}
\newcommand{\ee}{\end{equation}}

\def\bc{\begin{center}}
\def\ec{\end{center}}
\def\bea{\begin{eqnarray}}
\def\eea{\end{eqnarray}}

\graphicspath{{./Pictures/13Dec2021}}

\newcommand{\probP}{\text{I\kern-0.15em P}}

\def\multiset#1#2{\ensuremath{\left(\kern-.3em\left(\genfrac{}{}{0pt}{}{#1}{#2}\right)\kern-.3em\right)}}

\begin{document}

\newcommand{\rev}[1]{{#1}}
\newcommand{\revrev}[1]{{#1}}

\title{\large\bfseries Percolation and topological properties of temporal higher-order networks}

\author{Leonardo Di Gaetano}
\affiliation{Department of Network and Data Science, Central European University, 1100 Vienna, Austria}

\author{Federico Battiston}
\email{battistonf@ceu.edu}
\affiliation{Department of Network and Data Science, Central European University, 1100 Vienna, Austria}

\author{Michele Starnini}
\email{michele.starnini@gmail.com}
\affiliation{Departament de Fisica, Universitat Politecnica de Catalunya, Campus Nord, 08034 Barcelona, Spain}
\affiliation{CENTAI Institute, 10138 Turin, Italy}

\date{\today}

\begin{abstract}
\rev{Many complex systems that exhibit temporal non-pairwise interactions can be represented by means of generative higher-order network models. 
Here, we propose a hidden variables formalism to analytically characterize a general class of higher-order network models.}
We apply our framework to a temporal higher-order activity-driven model, providing analytical expressions for the main topological properties of the time-integrated hypergraphs, depending on the integration time and the activity distributions characterizing the model. 
Furthermore, we provide analytical estimates for the percolation times of general classes of uncorrelated and correlated hypergraphs.
Finally, 
we quantify the extent to which the percolation \rev{time} of empirical social interactions is underestimated when their higher-order nature is neglected.

\end{abstract}

\maketitle


An extremely broad category of complex systems can be represented as networks, where nodes describe units and links encode their pairwise interactions~\cite{boccaletti2006complex}. 
Despite widespread use, the dyadic structure does not allow for an accurate description of all those systems where non-pairwise interactions play a fundamental role, from human~\cite{benson2018simplicial} and animal~\cite{musciotto2022beyond} social networks to collaboration networks~\cite{patania2017shape}, drug recombination~\cite{zimmer2016prediction}, cellular networks~\cite{klamt2009hypergraphs}, species interactions~\cite{levine2017beyond} and the human brain~\cite{petri2014homological,giusti2016two, santoro2022unveiling}.
Such systems are better described by hypergraphs~\cite{berge1973graphs}, where hyperedges encode interactions among an arbitrary number of system units~\cite{battiston2020networks}.
Taking into account higher-order interactions has been shown to significantly affect collective behaviors in networked dynamics~\cite{battiston2020networks, battiston2021physics}, including diffusion~\cite{schaub2020random,carletti2020random}, synchronization~\cite{bick2016chaos,skardal2020higher,millan2020explosive,lucas2020multiorder,gambuzza2021stability,zhang2023higher}, contagion~\cite{iacopini2019simplicial,chowdhary2021simplicial,neuhauser2020multibody} and evolutionary~\cite{alvarez2021evolutionary, civilini2021evolutionary,civilini2023explosive} processes.

Furthermore, networks are inherently dynamic, with interactions evolving in time~\cite{temporalnetworksbook}.
While extensive research has been devoted to model temporal networks~\cite{Takaguchi:2012, perra2012activity, moinet2015burstiness} and the behavior of dynamical processes unfolding on their top~\cite{PhysRevE.83.025102, Scholtes:2014aa, citeulike:12739344, Moinet_2019}, the interest in temporal higher-order networks blossomed only recently. 
Higher-order interactions have been observed to occur in bursts in real face-to-face interaction systems~\cite{cencetti2021temporal} and display temporal correlations among different orders~\cite{gallo2023higher}, and temporal dynamics is known to affect the epidemic threshold in higher-order models of social contagion~\cite{chowdhary2021simplicial, st2021universal}. 
With a few notable exceptions~\cite{petri2018simplicial,gallo2023higher}, most models of higher-order networks are static, generalizations of Erdos-Renyi~\cite{barthelemy2022class} or configuration models~\cite{courtney2016generalized,young2017construction,chodrow2020configuration}, or are limited to networks which grow over time~\cite{kovalenko2021growing,millan2021local}.
Modeling temporal group dynamics and predicting their connectivity properties at the microscale is still an open problem.

Here, we introduce a general approach to analytically characterize higher-order time-varying networks by means of a hidden variables (HV) framework. 
In pairwise networks, HV were introduced to model the presence of links in networks with structural correlations~\cite{boguna2003class}.
\rev{Until now, the HV formalism has been employed across a vast spectrum of first-order generative processes, such as to map networks into embedded spaces, including latent~\cite{rastelli2016properties} and hyperbolic spaces~\cite{kitsak2020link}, fitness models~\cite{caldarelli2002scale, hoppe2014percolation}, protein interaction~\cite{miller2007clustering} and social distance~\cite{boguna2004models}. 
Furthermore, the HV formalism has been applied to networks evolving over time \cite{hartle2021dynamic}, networks with inherent correlations~\cite{boguna2003class}, and subsequently employed to pinpoint the topological characteristics of activity-driven networks~\cite{starnini2013topological,starnini2014temporal,moinet2015burstiness}.}
\rev{However, the aforementioned works neglected the higher-order organization of the considered social and biological systems.} 

\rev{In this letter, we propose a higher-order HV formalism that provides a powerful approach to describe higher-order networked systems, applicable to a wide range of generative models.}
\rev{As a demonstration of its versatile applicability, we apply our framework to a higher-order activity-driven model, where group interactions of different sizes are generated over time.}
We study the connectivity properties of the time-integrated system, obtaining analytical \rev{asymptotic} expressions for the hyper-degree distribution and hyper-degree correlations over time. \rev{We obtain these results in the limit of sparse networks and large hyper-degrees.}
We provide analytical estimates for the percolation times of general classes of uncorrelated and correlated hypergraphs marking the onset of a giant connected component in the higher-order systems.
We conclude by showing that neglecting the higher-order nature of interactions in empirical social networks leads to systematically underestimating the percolation threshold, with implications for any dynamical process running on such systems.


\emph{Higher-order hidden-variable formalism.}
We start by developing the HV formalism for higher-order networks.
Each node $i$ of a network of $N$ nodes is endowed with an intrinsic \rev{vectorial HV $\vec{h}_i = (h^{(1)}_i,h^{(2)}_i, \ldots h^{(m)}_i, \ldots)$,
where the HV $h^{(m)}_i$ determines the $m-$order interactions of node $i$. 
For each order $m$, $h^{(m)}_i$ is drawn from an independent distribution $\rho(h^{(m)})$.}
The higher-order HV model assumes that the existence of a $m$-order hyperlink ($m-$link) among $m+1$ nodes depends only on their HV, i.e., a connection probability \rev{$\probP(h_{1}^{(m)}, \ldots, h_{m}^{(m)},h_{m+1}^{(m)})$}.
\rev{In general terms,} the hyper-degree distribution $P(k^{(m)})$  (being $k^{(m)}$ the number of $m-$links of a node) can be written as a function of the HV distribution as
\begin{equation}
    P(k^{(m)}) = \sum_{h^{(m)}} g(k^{(m)}|h^{(m)})\rho(h^{(m)}) ,
    \label{eq:deg-distr-hv}
\end{equation}
where $g(k^{(m)}|h^{(m)})$ is the conditional probability (propagator) that a node with HV $h^{(m)}$ ends up with a certain hyper-degree $k^{(m)}$.

As in the first-order case \cite{boguna2003class}, the propagator can be expressed as the convolution of partial propagators. For instance, for $m=2$, 
\revrev{
\begin{equation}
    g(k^{(2)}|h^{(2)}) =  \sum_{ \{ k_{ij}^{(2)} \}} \delta_{\sum{k_{ij}^{(2)}}}^{k^{(2)} } \prod_{i\geq j}^C g^{(h^{(2)})}_{ij}(k^{(2)}_{ij}|h_i^{(2)}, h_j^{(2)} ),
    \label{eq:2-prop}
\end{equation} }
where $g^{(h^{(2)})}_{ij}(k_{ij}^{(2)}|h_i^{(2)}, h_j^{(2)})$ is the probability that a node (with HV $h^{(2)}$) ends up \rev{with $k_{ij}^{(2)}$ $2-$order interactions, with neighbors of HV $h_i^{(2)}$ and $h_j^{(2)}$.}
In the convolution, we take into account all the possible pairs of classes of HV excluding permutations ($i \geq j$), being
$h_C^{(2)}$ the maximum value of $h^{(2)}$ \revrev{and we sum over the set of all possible $2-$degree values $\{ k_{ij}^{(2)}\} = \{ k_{11}^{(2)}, k_{12}^{(2)} \ldots k_{CC}^{(2)} \}$}.
The Kronecker delta 
constrains that the final $2-$degree $k^{(2)}$ is equal to the sum of the partial degrees $k_{ij}^{(2)}$. See SM for the \revrev{explicit} $m$-order general expression.

For any $m$, one can solve the convolutional equation by resorting to the generating function of the propagator, $\rev{\hat{g}(z|h^{(m)})} = \sum_k z^{k^{(m)}} g(k^{(m)}|h^{(m)})$, \rev{where, for a lighther notation, from now on we indicate $h^{(m)}$ as $h$}.
Since the propagator is the convolution of partial propagators, given by Eq. \eqref{eq:2-prop}, its generating function is equal to the product of the generating functions of the partial propagators.
If hyperlinks are independently drawn according to the HV of nodes, the partial propagators are binomial distributions, and their generating functions can be obtained easily (see SM).
The logarithm of the generating function can be eventually written as a function of the HV distribution and the connection probability $\probP(h, h_{1},\ldots, h_{m})$,
\begin{multline}
    \ln{(\hat{g}(z|h)) }= \frac{N^m}{m!} \sum_{{h_1},\ldots, {h_m}} \rho(h_1) \ldots \rho(h_{m}) \\
    \ln{\bigg[1- (1-z)\probP(h, h_{1},\ldots, h_{m}) \bigg]},
    \label{eq:m-gener-funct}
\end{multline}
where one has to sum (integrate) over $m$ HV distributions and the factor $m!$ comes from excluding permutations. 

In the limit of sparse networks, $\probP(h, h_{1},\ldots, h_{m})  \ll 1$, the generating function of the propagator is exponential, thus indicating that the propagator is a Poisson distribution for every order $m$, as in the dyadic case $m=1$ \cite{boguna2003class}.
\rev{From the generating function of the propagator $\hat{g}$, one can compute the expected $m$-degree of a node with HV $h$ 
by means of the first derivative of $\hat{g}(z|h)$ at $z=1$ \cite{boguna2003class}, and it reads  }
\begin{equation}
\overline{k}^{(m)}(h)= \frac{N^m}{m!} \sum_{{h_{1}},\ldots, {h_{m}}}  \rho(h_{1}) \ldots \rho(h_{m}) \probP(h, h_{1}, \ldots, h_{m}).
 \label{eq:prop-mean}
\end{equation}
Instead, the problem-specific piece of information that allows us to treat different models is contained in Eq. \eqref{eq:prop-mean} through the connection probability $\probP(h, h_{1}, \ldots, h_{m})$, which is the key ingredient to find the hyper-degree distribution, given by Eq. \eqref{eq:deg-distr-hv}.

Similarly, we can study hyper-degree correlations starting from the conditional connection probability. 
We define the average $m-$degree of the neighbours of a node with HV $h$ as (see SM),
\begin{multline}
\overline{k}^{(m)}_{nn}(h) = \\
\sum_{h_{1},\ldots, h_{m}}  \bigg( \frac{ \overline{k}^{(m)}(h_{1})+\ldots + \overline{k}^{(m)}(h_{m})}{m} \bigg) p(h_{1}, \ldots, h_{m} |h),
    \label{eq:knn-h}
\end{multline}
where 
$p(h_1, \ldots, h_m|h)$ is the conditional probability that a node with HV $h$ is connected to nodes with HV $h_1, h_2 \ldots h_m$.
\revrev{The average $m-$degree of the neighbours of a node with $m-$degree $k$, }
$\overline{k}^{(m)}_{nn}(k)$, can be eventually found by following \cite{boguna2003class}, obtaining a form equivalent to the first-order case. 
Therefore, the HV formalism allows us to obtain the hyper-degree correlations of a large variety of higher-order generating processes simply by knowing the HV distribution and the connection probability depending on these variables.

\emph{The higher-order activity-driven (HOAD) model.}
We apply the higher-order HV framework to 
the higher-order activity-driven (HOAD) model, describing temporal group dynamics,
inspired by a very similar model for simplicial complexes~\cite{petri2018simplicial}. 
Each agent $i$ in a population of size $N$ is endowed with a higher-order activity potential $\textbf{a}_i = (a_i^{(1)}, a_i^{(2)}, ..., a_i^{(m)})$ for every interaction order $m$. 
The activities of the agents are random variables, extracted from distributions $\rho(\textbf{a}) = (\rho(a^{(1)}),\rho(a^{(2)}), ..., \rho(a^{(m)}))$, which we assume independent.
The {activity} of node $i$ at order $m$, $a_i^{(m)}$, represents the probability that they engage in an interaction with $m$ other nodes in a certain time-interval $\Delta t$.

The HOAD model generates temporal hypergraphs 
starting by $N$ initially disconnected nodes. 
At every time step, each node $i$ generates one hyperlink of order $m$ 
towards randomly selected nodes, with probability proportional to their activity $a_i^{(m)}$. 
At the following time step, the existent higher-order interactions are erased and the process continues. 
The temporal hypergraph is defined by the sequence of instantaneous, sparse hypergraphs generated at each time step. 
One can obtain a static hypergraph by integrating all instantaneous hypergraphs up to a certain time $T$, where two nodes $i$ and $j$ will be connected if any hyperedge between them exists in any instantaneous hypergraph in $t \in [1,T]$


\emph{Topological properties of HOAD networks.}
We now compute the topological properties of the HOAD networks integrated up to a certain time $T$, by mapping the HOAD model to the HV formalism.
First, since the activity distributions of different orders are assumed independent, we treat every order separately and omit the order dependency for brevity from now on, $a=a^{(m)}$.
The key step for the HV mapping resides in computing the probability that a node with activity potential ${a}_i$  will be connected with a set of $m$ other nodes, with activity ${a}_{1}, \ldots, {a}_{m}$, at time $T$ in the integrated network, namely $\probP_T(a,a_{1}, \ldots, a_{m})$.
By following \cite{starnini2013topological}, we can find this expression starting from the probability $\mathcal{Q}_T(a,a_{1}, \ldots, a_{m}) = 1- \probP_T(a,a_{1}, \ldots, a_{m})$ that the set of $m+1$ nodes is not connected by a $m$-link until time $T$.
Considering that every time a node is active it selects $m$ random neighbors for a $m$-link, and that the number of times a node can be active till time $T$ is described by a binomial distribution, we write 
\begin{equation}
    \probP_T(a,a_{1}, \ldots, a_{m}) \simeq \frac{ m! }{N^m}(a+a_{1}+\ldots+a_{m}) T,
    \label{eq:conn-prob}
\end{equation}
 where we have worked in the limit of $N \gg T \gg 1$ (see SM).
By inserting Eq. \eqref{eq:conn-prob} into Eq. \eqref{eq:prop-mean}, we can obtain for $N \gg m$ the expected $m-$degree of nodes with activity $a$, at time $T$, 
\begin{equation}
 \overline{k}^{(m)}_T(a) \simeq T(a + m\langle a\rangle),
 \label{eq:mean-degree}
\end{equation}
where $\langle a \rangle = \sum_a a \rho(a)$ denotes the usual average of activity of order $m$ over the population.
The expected $m-$degree is intuitively equal to $a T$ outgoing $m$-links plus $m T \langle a \rangle$ connections received from random neighbors.

\begin{figure}[tbp]
    \centering
    \includegraphics[width =\columnwidth]{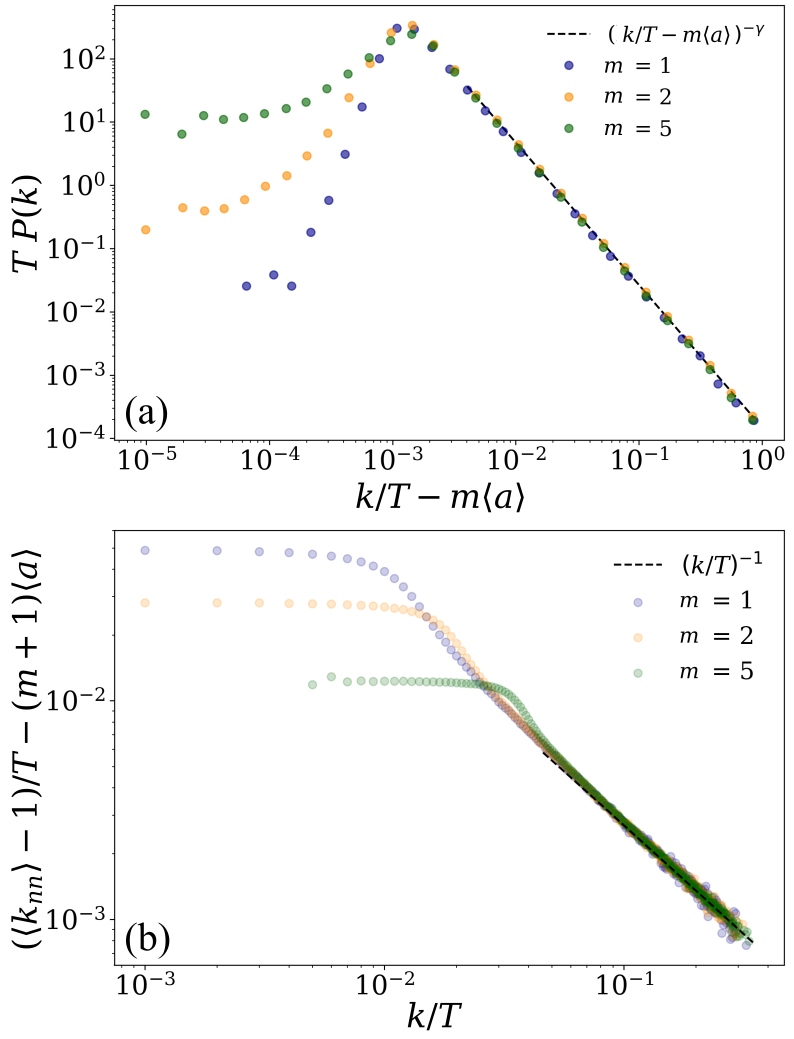}
    \caption{ \textbf{Topological properties of HOAD networks.} (a) Hyper-degree distribution $P_T(k^{(m)})$,     Eq. \eqref{eq:deg-distr-ad} shown as a dashed line.
    (b) Hyper-degree correlations $\overline{k}^{(m)}_{nn,T}(k)$,
    Eq. \eqref{knn last equation} shown as a dashed line. Network size $N = 10^6$, orders $m=1,2,5$, integration time $T=10^3$. 
    Different values of $T$ and $m$ are shown in SM. The activity distributions $\rho(a)$ of order $m$ have power-law form for every order with exponent $\gamma = 2.25$.  
        }
    \label{fig:Fig12}
\end{figure}

By inserting Eq. \eqref{eq:mean-degree} into the Poissonian form of the propagator 
and substituting it into Eq. \eqref{eq:deg-distr-hv}, one can obtain the asymptotic limit of the degree distribution of order $m$ of the aggregated network till time $T$ (see SM),
\begin{equation}
 P_T(k^{(m)})  \simeq  \frac{1}{T}\rho \bigg(\ \frac{k^{(m)} }{T} -m\langle a\rangle \ \bigg).
 \label{eq:deg-distr-ad}
\end{equation}
The last expression is obtained in the limit of large $N$ and small $T$, $T \ll \frac{N^m}{m!}$ and for $T^2 \gg k^{(m)} \gg 1$, (see SM).
Figure \ref{fig:Fig12} (a) shows the hyper-degree distribution $P_T(k^{(m)})$ of HOAD networks integrated at time $T$, as obtained by numerical simulations. 
\rev{We arbitrarily select a power-law activity distribution, yet Eq. \eqref{eq:deg-distr-ad} is general for any distribution $\rho$.}
The model is implemented as part of the library HGX~\cite{lotito2023hypergraphx}.
One can see a good agreement with the asymptotic behavior indicated by Eq. \eqref{eq:deg-distr-ad}.

\emph{Hyper-degree correlations of HOAD networks.}
The average $m-$degree of the neighbors of a node with activity $a$ at time $T$, $\overline{k}^{(m)}_{nn,T}(a)$, is obtained by the HV mapping of Eq. \eqref{eq:knn-h}.
To this aim, one needs to compute the conditional probability $p(a_{1}, a_{2}\ldots, a_m|a)$ that a node with activity $a$ is connected to nodes with activities $a_{1}, a_{2}\ldots, a_m$, by using the connection probability of the HOAD model, given by Eq. \eqref{eq:conn-prob}, and the expected $m-$degree of nodes with activity $a$ at time $T$, given by Eq. \eqref{eq:mean-degree}.
After obtaining $\overline{k}^{(m)}_{nn,T}(a)$ (see SM for the analytical expression), the $m-$order degree-degree correlation can be obtained by following \cite{boguna2003class} and it reads
\begin{equation}
    \frac{\overline{k}^{(m)}_{nn,T}(k) -1}{T} \simeq (m+1) \langle a \rangle + \sigma^2 \bigg(\frac{k^{(m)}}{T}\bigg)^{-1},
    \label{knn last equation}
\end{equation}
where  $\sigma^2 = \langle a^2\rangle - \langle a\rangle^2$ of the $m-$order activity.


The last expression, valid in the limit of $k^{(m)} \gg 1$ and sparse network (SM), gives an asymptotic prediction of $\overline{k}^{(m)}_{nn,T}(k)$ as a function of the first two momenta of the activity distribution of order $m$. 
Figure \ref{fig:Fig12} (b)  shows 
\rev{the correlations minus its first moment}
of HOAD networks integrated at time $T$, as obtained by numerical simulations. 
As for the degree distribution, we plot the rescaled hyper-degree correlations, \rev{the differences between the correlations and their leading approximation} in order to show \rev{how it decays with \( T/k^{(m)} \)} and the collapse of the curves for three different orders $m=1,2,5$.
One can see that the disassortative behavior proportional to $(k^{(m)})^{-1}$ and governed by $\sigma^2$, as predicted by Eq. \eqref{knn last equation}, is confirmed by numerical simulations.

\emph{Temporal percolation of HOAD networks.}
The connectivity properties of the time-integrated HOAD networks allow us to characterize the temporal percolation, i.e., the time $T_p$ marking the onset of a giant connected component in the integrated network.
The percolation time $T_p$ is particularly relevant for dynamical processes unfolding of these temporal networks, since any process with a characteristic lifetime smaller than $T_p$ will be unable to explore a sizable fraction of the network.

\rev{
The details of the derivation of the percolation times are reported in the Appendix. 
We first obtain the conditions for the percolation threshold of static correlated and uncorrelated hypergraphs of order $m$.
Then, we map these results into the HOAD model, by writing the degree momenta as a function of the activity distribution, thus finding the percolation times for correlated and uncorrelated HOAD networks. 
} 
The percolation time for correlated HOAD networks of order $m$ reads
\begin{equation}
    T^{(m)}_c = \frac{2}{\langle a \rangle (m+1) + \sqrt{\langle a \rangle^2 (m^2+2m-3)+4\langle a^2 \rangle} }.
    \label{correlated percolation last equation}
\end{equation}

\begin{figure}[tbp]
    \centering
    \includegraphics[width =\columnwidth]{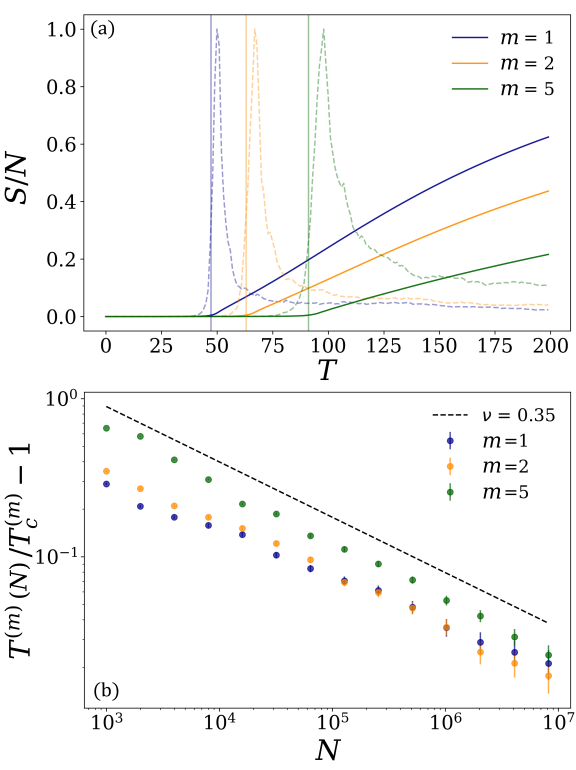}
    \caption{\textbf{Percolation time of HOAD networks.} Orders $m = 1, 2, 5$. 
    (a) Giant component size $S/N$ (continuous line) and the peak of its variance $\sigma(S)^2$ (dashed line) over time. 
    The theoretical prediction given by Eq. \eqref{correlated percolation last equation} is indicated as a vertical line.
    (b) Finite-size scaling analysis of the relative difference $(T^{(m)}(N) - T_c^{(m)})/T_c^{(m)}$ (circles) and corresponding scaling law $N^{-\nu}$ (dashed line).
    Results are averaged over $10^2$ runs.}
    \label{fig: Percolation figures}
\end{figure}

We test the validity of the prediction given by Eq. \eqref{correlated percolation last equation} by running extensive numerical simulations. 
Figure \ref{fig: Percolation figures} (a) shows the growth of the giant component size $S$ over time and the peak of its variance, $\sigma(S)^2$,  indicating the estimated percolation \rev{time}, for several orders $m$. 
The percolation \rev{time} predicted by Eq. \eqref{correlated percolation last equation} has a decent agreement with numerical results, yet they do not exactly coincide.
We thus run a finite-size scaling analysis, by assuming that the relative difference between the actual percolation \rev{time} $T_{c}^{(m)}$ in the thermodynamic limit and the one found in a network of size $N$, $T^{(m)}(N)$, follows a scaling law of the form $(T^{(m)}(N) - T_c^{(m)})/T_c^{(m)} \sim N^{-\nu}$ for every $m$.
Figure \ref{fig: Percolation figures} (b) shows that the finite-size hypothesis holds, the percolation \rev{time} estimated by the peak over time of the variance of the giant component size actually approaches $T_{c}^{(m)}$ for any order $m$ in the thermodynamic limit $N \to \infty$.


\begin{figure}[tbp]
    \centering
    \includegraphics[width =\columnwidth]{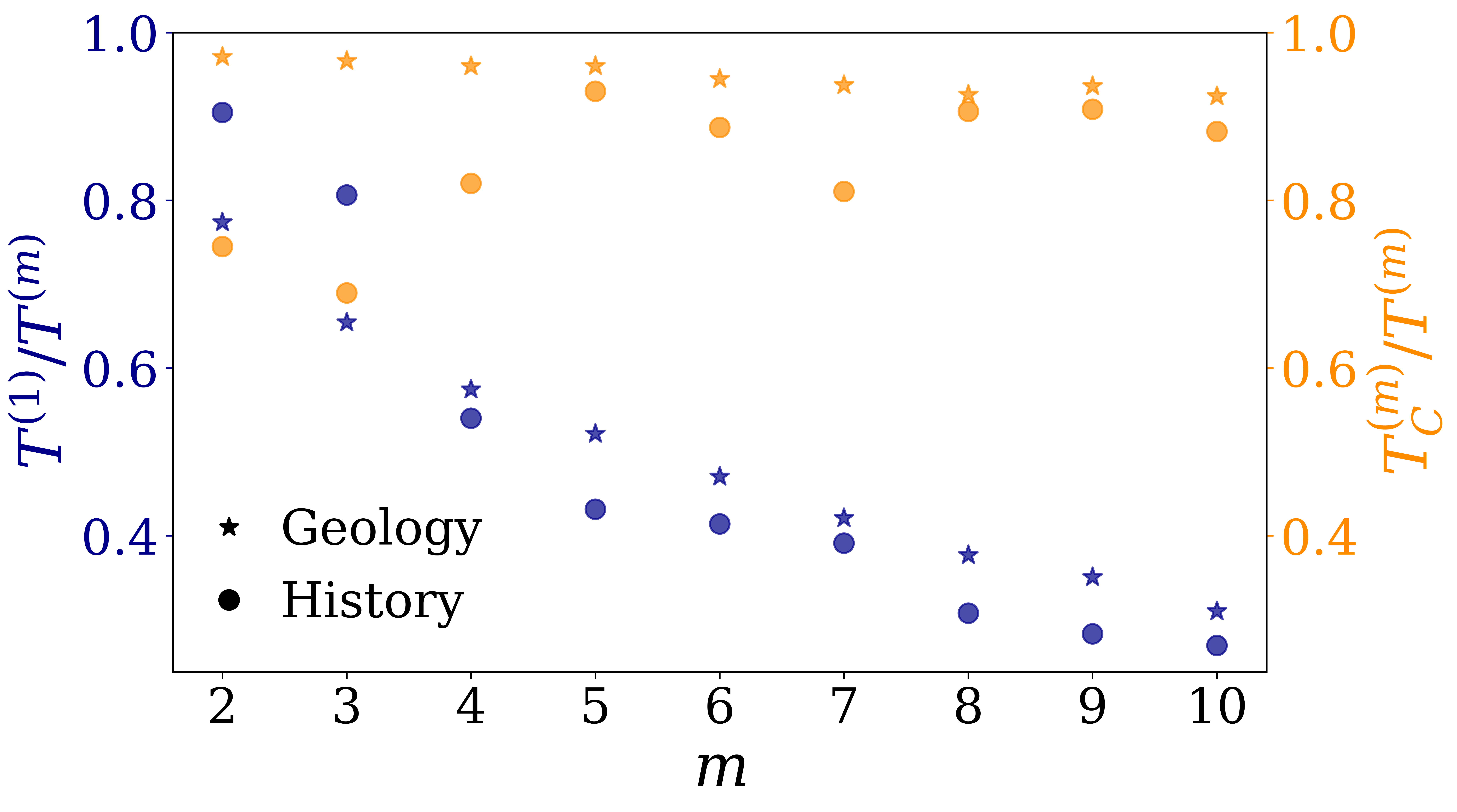}
    \caption{\textbf{Percolation \rev{times} in empirical data.} 
    Scientific Geology (\rev{stars}) and History (\rev{circles}) collaboration networks.
    Blue points: Ratios between the first-order ($T^{(1)}$) and $m-$order ($T^{(m)}$) percolation times of networks informed by empirical activities, estimated from numerical simulations.
    Yellow points: Ratios between the theoretical prediction from Eq. \eqref{correlated percolation last equation}, $T_c^{(m)}$, and the percolation times of networks informed by empirical activities estimated from numerical simulations, $T^{(m)}$.}
    \label{fig:data figure}
\end{figure}

\emph{Empirical data.}
Finally, we show the potential of the HOAD modeling framework by testing the theoretical predictions for the percolation time on higher-order empirical data, comparing it with the first-order percolation. 
For this latter case, we project all interactions into the first order, thus representing  higher-order data as a simple network, losing part of the information contained therein. 
We consider two data sets of scientific collaboration networks in the fields of Geology and History, collected by the Microsoft Academic Graph (see SM for details). 
We inform first-order and higher-order activity-driven models with empirical activities extracted from the dataset, and compare the first-order ($T^{(1)}$) and $m-$order ($T^{(m)}$) percolation times of the networks.
The percolation \rev{points} are obtained by calculating the time for which the variance of the component sizes distribution is maximum.

Figure \ref{fig:data figure} shows that the $m-$order percolation time $T^{(m)}$ estimated by numerical simulations of the HOAD model informed by empirical data is in good agreement with the theoretical prediction $T_c^{(m)}$ given by Eq. \eqref{correlated percolation last equation}, for every order $m$.
Moreover, Figure \ref{fig:data figure} shows that the first-order percolation time $T^{(1)}$ is much smaller than the actual $m-$order one $T^{(m)}$, and such a difference increases with the order $m$.
Therefore, an incorrect representation of higher-order data as classic, dyadic interactions leads to a substantial underestimation of the true, higher-order percolation \rev{times}, up to $50\%$ already for $m=5$, that is, small groups of 6 people.  


\emph{Conclusions.}
In this work, we showed that the topological and percolation properties of temporal higher-order networks can be obtained by mapping such networks to a higher-order HV formalism. 
We illustrate the potential of our theoretical framework by quantitatively showing how much the percolation times of higher-order empirical social networks are underestimated if higher-order interactions are neglected. 
\rev{This result is particularly interesting within the framework of epidemic processes: a disease spreading with a short timescale is expected to percolate when the underlying contact network is assumed to be formed by dyadic interactions, but it would not percolate in the corresponding higher-order network representation. 
Note, however, that our finding holds within the specific activity-driven modeling framework. 
Further research should be devoted to addressing this setting in different modeling frameworks and on real contact networks.}

\rev{The higher-order HV framework we developed holds potential for future applications across a wide array of higher-order and temporal generative models.
For instance, it could be applied to higher-order fitness models
~\cite{caldarelli2002scale} or social dynamics models including higher-order interactions mapped into latent spaces~\cite{boguna2004models}.
Likewise, it could be extended to describe network models incorporating  Non-Markovian dynamics~\cite{moinet2015burstiness}, which has shown to have a deep impact on epidemic processes.}
Future research could quantify and model the presence of correlations between different interaction orders, as well as their effects on the connectivity and percolation properties of time-integrated networks.
We hope that our work will stimulate further research to apply the higher-order HV framework to other empirical, time-varying complex systems. 

\section*{Acknowledgments}
We acknowledge Romualdo Pastor-Satorras for useful discussions. F.B. acknowledges support from the Air Force Office of Scientific Research under award number FA8655-22-1-7025.

\appendix
\section*
{Appendix on Temporal Percolation}
\setcounter{equation}{0}
\renewcommand{\theequation}{A.\arabic{equation}}
Here we show how we calculate the percolation times for correlated and uncorrelated HOAD networks. 
We first consider static hypergraphs of order $m$ 
whose nodes may be removed with probability $1 - p$. 
Following the approach outlined in \cite{goltsev2008percolation}, we determine the probability $x_k$ of avoiding a giant connected component while traversing a $m$-order hyperlink (connecting $m+1$ nodes) starting from a node with hyper-degree $k$ (we omit the dependency in $m$). 
This condition can be expressed as 
\begin{equation}
x_{k} = \bigg[ 1- p + p  \sum_{k'} P(k'| k) x_{k'}^{k'-1}\bigg]^m,
\label{eq:x_k eq}
\end{equation}
 where $P(k'| k)$ is the probability that a node with $m$-degree $k$ is connected with a node of $m$-degree $k'$ and we assume that the probability $x_{k'}$ of each of the $m$ nodes to be connected to the giant component is independent of each other, so we exponentiate the same probability to the $m$.
 Close to the percolation threshold we have $x_k \lessapprox  1$, hence defining $y_k = 1 - x_k \gtrapprox 0$ 
and expanding \rev{Eq. \eqref{eq:x_k eq} (see SM for detailed calculations)}, 
we get
\begin{equation}
    y_k = mp \sum_{k'} \textbf{B}^{(m)}_{kk'} y_{k'}, 
    \label{eq: branching matrix}
\end{equation}
\rev{where we have defined the $m$-order branching matrix as $\sum_{k'} \textbf{B}^{(m)}_{kk'} y_{k'} = \sum_{k'} (k'-1) P(k'| k) y_{k'} $.}
\rev{Besides a multiplicative factor $m$, Eq. \eqref{eq: branching matrix} is equivalent to the result found for simple networks \cite{goltsev2008percolation}, having also the same element-wise representation of $ \textbf{B}^{(m)}_{kk'} $ for every $m$ (see SM).}
The last expression also allows us to find the percolation condition for uncorrelated hypergraphs, by writing the conditional probability as $P(k'| k) = {\rho(k')k'}/{\langle k \rangle}$. 
In this way, we find the $m$-order version of the well-known Molloy-Reed criterion \cite{molloy1995critical}, ${\langle k^2 \rangle- \langle k \rangle}/{\langle k \rangle} > {1}/{m}$, already found in \cite{sun2021higher}. 
By explicitly writing the degree momenta as a function of the activity distribution, the percolation time for uncorrelated hypergraphs reads
\begin{equation}
    T_{unc}^{(m)}=\frac{\left(m + 1\right)\langle a \rangle }{m(m +2) \langle a \rangle^{2}  + \langle a^2 \rangle}.
\end{equation}

The percolation threshold for correlated networks is instead given by the condition $m p_c\lambda_1^{(m)} = 1$ from Eq. \eqref{eq: branching matrix}, where $\lambda_1^{(m)}$ is the dominant eigenvalue of the $m$-order branching matrix $\textbf{B}^{(m)}_{kk'}$, as guaranteed by the Perron-Froebenius theorem \cite{goltsev2008percolation}, and $p_c$ is the critical density of nodes for the onset of a giant connected component. 
The largest eigenvalue $\lambda_1^{(m)}$ can be found by means of the HV formalism, by following \cite{starnini2014temporal}, as a function of the first and second degree momenta of order $m$, $\langle k \rangle_{T_p}$ and $\langle k^2 \rangle_{T_p}$ (see SM).
We then map these expressions into the HOAD model, where the degree momenta are given by the activity distributions, and find the percolation \rev{time} for correlated HOAD networks, Eq. \eqref{correlated percolation last equation} of the main text.

Both analytical predictions for correlated and uncorrelated networks depend on the first two momenta of $\rho(a)$.
For large $m$, we have $T_{c,unc}^{(m)} \propto \frac{1}{m} \rightarrow 0$ for both correlated and uncorrelated cases, so the uncorrelated \rev{percolation time} approaches the correlated one in this limit. 
The difference between the two \rev{times} is maximum for strongly heterogeneous networks, see SM.


%

\newpage
\onecolumngrid
\vspace{\columnsep}

\section{Hidden variables formalism for higher-order networks}

Hidden variable models {for complex networks} are based on two assumptions: \revrev{\textit{i)}} each node $i$ has a hidden variable \revrev{$ \vec{h}_i$, with components $h^{(m)}_i$ for every order of interaction $m$,} drawn from a probability distribution $\rho(h^{(m)})$, and \revrev{\textit{ii)}} the probability that a set of $m+1$ nodes $[1,2, \ldots m+1]$ of hidden variables $[h^{(m)}_1,h^{(m)}_2, \ldots h^{(m)}_{m+1}]$ belongs to the same $m$-order link depends only on such hidden variables, $\probP(h^{(m)}_1,h^{(m)}_2, \ldots h^{(m)}_{m+1})$.

\subsection{Hyper-degree distribution}

The $m$-order degree distribution $P(k^{(m)})$ can be written as 
\begin{equation}
    P(k^{(m)}) = \sum_{h^{(m)}} g(k^{(m)}|h^{(m)})\rho(h^{(m)}),
    \label{degree distribution m order}
\end{equation}
where the propagator $g(k^{(m)}|h^{(m)})$ is the probability that a node with hidden variable $h^{(m)}$ ends with a $m$-order degree equal to $k^{(m)}$, i.e., it has $k^{(m)}$ incident $m$-links. 
Note that $\sum_{k^{(m)}} g(k^{(m)}|h^{(m)}) = 1$.
As it has been done for pairwise networks \cite{boguna2003class}, we can express the propagator as the convolution of all possible conditional probabilities that lead to it, namely the partial propagators.
For 
the $2$-order case:
\begin{equation}
    g(k^{(2)}|h^{(2)}) =  \sum_{ 
    k^{(2)}_{11},k^{(2)}_{12}, \ldots k^{(2)}_{CC} }
\delta_{ k^{(2)}_{11}+k^{(2)}_{12}+ \ldots +k^{(2)}_{CC}}^{k^{(2)}} \   g^{(h^{(2)})}_{11}(k^{(2)}_{11}|h^{(2)}_1, h^{(2)}_1 ) \ g^{(h^{(2)})}_{12}(k^{(2)}_{12}|h^{(2)}_1, h^{(2)}_2 )\ldots g^{(h^{(2)})}_{CC}(k^{(2)}_{CC}|h^{(2)}_C, h^{(2)}_C ),
\nonumber
    \label{eq:2-prop}
\end{equation}
shortly,
\begin{equation}
    g(k^{(2)}|h^{(2)}) =  \sum_{ \{k^{(2)}_{ij} \}} \delta_{\sum{k^{(2)}_{ij}}}^{k^{(2)}}  \prod_{i\geq j}^C g^{(h^{(2)})}_{ij}(k^{(2)}_{ij}|h^{(2)}_i, h^{(2)}_j ),
    \label{eq:2-prop}
\end{equation}
where $g^{(h^{(2)})}_{ij}(k^{(2)}_{ij}|h^{(2)}_i, h^{(2)}_j )$ is the probability that a node (with hidden variable $h^{(2)}$) ends up with a number of $2$-order interactions with neighbors of hidden variables $h^{(2)}_i$ and $h^{(2)}_j$ equal to $k^{(2)}_{ij}$.
In the convolution, we take into account all the possible pairs of classes of hidden variables excluding permutations ($i \geq j$), being $h^{(2)}_C$ the maximum value of $h^{(2)}$ \revrev{and we sum over the set of all possible $2-$degree values $\{ k_{ij}^{(2)}\} = \{ k_{11}^{(2)}, k_{12}^{(2)} \ldots k_{CC}^{(2)} \}$}.
Note that the number of all possible partial propagators is equal to the number of multisets of cardinality 2 among $C$ elements, $\multiset{C}{2} = {C+2-1 \choose 2}$.
The term $\delta_{\sum{k_{ij}}}^{k^{(2)} } $ constraints the sum of partial degrees to be equal to $k^{(2)}$.
\revrev{For simplicity, from now on we omit the explicit dependence of $h^{(m)}$ in $m$  without losing generality in the discussion.}

For the $m$-order case, one has to consider that a $m$-order interaction between the node with hidden variable $h$  and other $m$ nodes involves $m$ (not necessarily different) hidden variable classes, $h_{i_1},h_{i_2}, \ldots, h_{i_m}$. \rev{Notice that we have dropped the explicit dependence of $h$ on $m$ to have a lighter notation, $h = h^{(m)}$.}
The propagator thus reads
\revrev{
\begin{equation}
    g(k^{(m)}|h) = \sum_{ \{k_{i_1i_2 \ldots i_m} \}} \delta_{\sum{ k_{i_1i_2 \ldots i_m}}}^{k^{(m)}} \prod_{i_1 \geq i_2 \geq \ldots \geq i_m }^C g^{(h)}_{i_1i_2 \ldots i_m}(k_{i_1i_2 \ldots i_m}|h_{i_1},h_{i_2}, \ldots, h_{i_m} ),
    \label{eq:m-prop}
\end{equation} }
As in the $m=2$ case, $g^{(h)}_{i_1i_2 \ldots, i_m}(k_{i_1i_2 \ldots i_m}|h_{i_1},h_{i_2}, \ldots, h_{i_m} )$ is the probability that a node with hidden variable $h$ has exactly $k_{i_1i_2 \ldots i_m}$  $m$-order interactions with neighbours of hidden variables $h_{i_1},h_{i_2}, \ldots, h_{i_m}$. 
In this case, the convolution is done again considering $i_1 \geq i_2 \geq \ldots \geq i_m $ in order to avoid all repetitions given by the permutation of the indexes. 
As for the second-order case, $C$ is the number of hidden variable classes.
The number of partial propagators in the convolution is $\multiset{C}{m} = {C+m-1 \choose m}$, namely the number of multisets with $m$ possibly repeated items, chosen from a set of $C$ distinct elements.
The term $\delta_{\sum{k_{i_1i_2 \ldots i_m}}}^{k^{(m)} } $ constraints the sum of partial degrees to be equal to $k^{(m)}$, and
\revrev{$\{k_{i_1i_2 \ldots i_m} \}$ is again the set of all possible values of $m-$degree.}

For the purpose of solving the convolution in Eq. \eqref{eq:m-prop}, we resort to the properties of generating functions.
The generating function of the propagator is defined as
\begin{equation}
    \hat{g}(z|h) = \sum_{k} z^{k} g(k|h),
\end{equation}
where we omit the $m$ index on the hyperdegree $k^{(m)}$.
Since the propagator is given by a convolution of Eq. \eqref{eq:m-prop}, we can write its generating function as the product of the generating functions of the partial propagators. 
For 
a general order $m$ we have 
\begin{equation}
    \hat{g}(z|h) = \prod_{i_1 \geq i_2 \geq \ldots \geq i_m } \hat{g}^{(h)}_{i_1,i_2, \ldots, i_m}(k_{i_1,i_2, \ldots, i_m}|h_{i_1},h_{i_2}, \ldots, h_{i_m} ) .
    \label{eq:gen-fun-con-m}
\end{equation}
Since the $m$-links between vertices with hidden variables $h, h_{i_1},h_{i_2}, \ldots, h_{i_m}$ are independently drawn with probability $\probP(h, h_{i_1},h_{i_2}, \ldots h_{i_m})$, the  partial propagator \\
${g}^{(h)}_{i_1,i_2, \ldots, i_m}(k_{i_1,i_2, \ldots, i_m}|h_{i_1},h_{i_2}, \ldots, h_{i_m} )$ is simply given by a binomial distribution, as in the first order case.
Consequently, its generating function reads
\begin{equation}
    \hat{g}^{(h)}(z|h_{i_1}, h_{i_2}, \ldots, h_{i_m}) = \big[1 - (1-z) \ \probP(h, h_{i_1}, h_{i_2}, \ldots, h_{i_m}) \big]^{N_{i_1i_2 \ldots i_m}},
    \label{generating function of partial propagator}
\end{equation}
where $N_{i_1,i_2 \ldots i_m}$ is the number of possible sets with nodes of hidden variables $h_{i_1}, h_{i_2}, \ldots, h_{i_m}$, that can be written as $N_{i_1i_2 \ldots i_m} = N_{i_1} N_{i_2} \ldots N_{i_m}$, where $N_{i_1}= N\rho(h_{i_1})$ is the number of nodes with hidden variable $h_{i_1}$. 
By taking the logarithm of the full propagator, one obtains
\begin{equation}
    \ln \hat{g}(z|h) = N^m \sum_{i_1 \geq i_2 \geq \ldots \geq i_m } \rho(h_{i_1}) \rho(h_{i_2}) \ldots \rho(h_{i_m}) \ln [1- (1-z)\probP(h, h_{i_1}, h_{i_2}, \ldots, h_{i_m})].
    \label{eq:log-gen-func}
\end{equation}
In the limit $C \gg m$, the number of elements in the summation, equal to the number of multisets $\multiset{C}{m}$, is equal to $\frac{C^m}{m!}$. We can thus sum over $m$ independent indexes $i_1, i_2 \ldots i_m = 1,2, \ldots C$, and divide by $m!$. 
At this point, since Eq. \eqref{eq:log-gen-func} does not depend anymore on the specific indexes $i_1, i_2 \ldots, i_m$, we can simplify the notation and directly sum over different hidden variable classes $h_1, h_2 \ldots, h_m$, 
\begin{equation}
    \ln \hat{g}(z|h) = \frac{N^m}{m!} \sum_{h_1, h_2 \ldots, h_m} \rho(h_{1}) \rho(h_{2}) \ldots \rho(h_{m}) \ln [1- (1-z)\probP(h, h_{1}, h_{2}, \ldots, h_{m})].
\end{equation}

We now consider the limit of sparse networks $N \gg 1$ and small connection probability $\probP(h, h_{1}, h_{2}, \ldots, h_{m}) \ll 1$, which allows us to write $\hat{g}(z|h)$ as a pure exponential generating function, as in the first order case. 
Consequently, the propagator takes the form of a Poisson distribution
\begin{equation}
    g(k^{(m)}|h) \simeq \frac{e^{-\overline{k^{(m)}}(h)}\overline{k^{(m)}}(h)^{k^{(m)}}}{k^{(m)}!},
    \label{m-poisson-propagator}
\end{equation}
where $\overline{k^{(m)}}(h)$ is the expected $m$-degree of a node of hidden variable $h$, that can be obtained by taking the first derivative of $\hat{g}(z|h)$ evaluated at $z=1$,
\begin{equation}
 \overline{k^{(m)}}(h)= \frac{N^m}{m!} \sum_{{h_{1}},\ldots, {h_{m}}}  \rho(h_{1}) \ldots \rho(h_{m}) \probP(h, h_{1}, \ldots, h_{m}). 
 \label{eq:prop-mean-approx}
\end{equation}

By inserting the form of the propagator Eq. \eqref{m-poisson-propagator} and its mean Eq. \eqref{eq:prop-mean-approx} into the general Eq. \eqref{degree distribution m order}, one can obtain the $m$-degree distribution as a function of the hidden variable distribution.
 The form of the propagator is exponential (as in the first order case) and the value of its mean depends explicitly on the connection probability.
 In the next section, we will discuss how to find $ \probP(h, h_{1}, \ldots, h_{m})$ in the case of a higher-order activity driven model.

\subsection{Hyper-degree correlations}


We now obtain {general} analytical expressions for the hyper-degree correlations {in higher-order networks with hidden variables}.
We start by indicating the average $m-$degree of the nearest neighbors of a node with hidden variable $h$. 
For $m=2$, one has to average over all possible hidden variables $h_i$ and $h_j$ of the two neighbors $i$ and $j$ in the $2-$link,
\begin{equation}
  \overline{k^{(2)}_{nn}}(h) = \sum_{h_i, h_j} \bigg( \frac{\overline{k^{(2)}}(h_i)+ \overline{k^{(2)}}(h_j)}{2} \bigg) p(h_i , h_j|h),
\end{equation}
where $p(h_i, h_j |h)$ is the conditional probability that a node with hidden variable $h$ is connected to nodes with hidden variables $h_i , h_j$.
Such conditional probability  can be written as 
\begin{equation}
      p(h_i, h_j |h) = \frac{N^2\rho(h_i)\rho(h_j)\probP(h,h_i, h_j)}{ 2 \ \overline{k^{(2)}}(h)},  
\end{equation}
where $N^2 \rho(h_i) \rho(h_j)$ is the total number of all possible pairs made up of one node with hidden variable $h_i$ and one with $h_j$,  $N^2 \rho(h_i) \rho(h_j) \probP(h,h_i,h_j)$ represents the expected number of $2$-links that a node with hidden variable $h$ shares with this type of pair, and the factor $2$ at the denominator ensures that the probability is correctly normalized when we sum over independent indexes, $\sum_{h_i,h_j}p(h_i, h_j | h)=1$ . 
For general $m$, the average $m$-degree of neighbors of a node with hidden variable $h$ reads 
\begin{equation}
 \overline{k^{(m)}_{nn}}(h) = \sum_{h_{1},h_{2}\ldots h_{m}}  \bigg( \frac{ \overline{k^{(m)}}(h_{1})+\overline{k^{(m)}}(h_{2})\ldots + \overline{k^{(m)}}(h_{m})}{m} \bigg) p(h_{1}, h_{2}\ldots h_{m} |h),
    \label{knn_h}
\end{equation}
 where $p(h_{1}, h_{2}\ldots h_{m} |h)$ is the conditional probability that a node with hidden variable $h$ is connected in a $m$-links with neighbours $h_{1}, h_{2}\ldots h_{m}$, which reads 
\begin{equation}
      p(h_1, h_2 \ldots , h_m|h) = \frac{N^m\rho(h_1)\rho(h_2)\ldots\rho(h_m) \probP(h,h_1, h_2 \ldots , h_m)}{ m! \ \overline{k^{(m)}}(h)},
      \label{eq:p-neigh}
\end{equation}
where, again, the correct normalization over independent indexes  is ensured by the term $m!$ leading to $\sum_{h_1, h_2 \ldots h_m}p(h_1, h_2 \ldots, h_m|h) = 1$.
The average $m$-degree of the nearest neighbors of a node with degree $k^m$ can be obtained by following Ref. \cite{boguna2003class}, 
\begin{equation}
    \overline{k^{(m)}_{nn}}(k^{(m)}) = 1+ \frac{1}{P(k^{(m)})} \sum_{h} g(k^{(m)}|h)\rho(h)\overline{k^{(m)}_{nn}}(h).
    \label{knn as knn_a}
\end{equation}
equivalently to the first-order case.

\section{The higher-order activity-driven model}

The higher-order activity-driven model (HOAD model) is defined as follows.
Each agent $i$ in a population of size $N$ is endowed with a higher-order activity potential $\textbf{a}_i = (a_i^{(1)}, a_i^{(2)}, ..., a_i^{(m)})$ for every interaction order $m$. 
The activities of the agents are random variables, extracted from distributions $\rho(\textbf{a}) = (\rho(a^{(1)}),\rho(a^{(2)}), ..., \rho(a^{(m)}))$, which we assume independent.
The {activity} of node $i$ at order $m$, $a_i^{(m)}$, represents the probability that they engage in an interaction with $m$ other nodes in a certain time-interval $\Delta t$.
The activity potentials can be measured in empirical data by considering that the activity $a_i^{(m)}$ is proportional to $n_i^{(m)}$, the number of interactions of order $m$ involving node $i$ in $\Delta t$. 
The proper normalization of $a_i^{(m)}$, $\sum_{i,m} a_i^{(m)} = 1$, implies that $a_i^{(m)}$ is equal to the number of interactions of order $m$ involving node $i$ in $\Delta t$, divided by the total number of interactions of any order all nodes are involved in $\Delta t$, $a_i^{(m)} = {n_i^{(m)}}/{\sum_{i,m} n_i^{(m)}}$.

We then map the HOAD model to the higher-order hidden variable formalism.
A node $i$ of the HOAD network with activity $\textbf{a}_i$ can be mapped as $\textbf{a}_i \rightarrow \vec{h}_i$.
Since hyperlinks of different orders are generated independently, we can treat all orders separately by means of $m$ distinct scalar hidden variables.
For every $m$ we can write $a_i^{(m)} \rightarrow h_i$.

\subsection{Hyper-degree distribution}

We now derive the analytical form of the distribution $ P(k^{(m)})$ by means of the hidden variables formalism. 
For the sake of simplicity, we first focus on the second-order case and compute the $\probP_T(a_i,a_j,a_k)$ that three nodes $i,j,k$ with hidden variables, $a_i$, $a_j$, and $a_k$ are connected by at least one $2$-link in the aggregated HOAD network at time $T$. 
By following \cite{starnini2013topological}, we start from the probability that these nodes are not connected, $\mathcal{Q}_T(a_i,a_j,a_k) = 1 - \probP_T(a_i,a_j,a_k)$.
Let $n_i,n_j, n_k$ be the number of activations of the three nodes until time $T$.
Since every time a node is active it selects two random neighbors, we have
\begin{equation}
   \mathcal{Q}_T(a_i,a_j,a_k) = \sum_{n_i, n_j, n_k} \rho_T(n_i) \rho_T(n_j) \rho_T(n_k) \bigg(1-\frac{1}{{N \choose 2}} \bigg)^{n_i} \bigg(1-\frac{1}{{N \choose 2}} \bigg)^{n_j} \bigg(1-\frac{1}{{N \choose 2}} \bigg)^{n_k},
    \label{Q integral}
\end{equation}
 where $ \rho_T(n_i)$ is the probability that node $i$ has been activated $n_i$ times at time $T$, given by a binomial distribution
\begin{equation}
    \rho_T(n_i) = {{TN \choose n_i}} \bigg( \frac{a_i}{N}\bigg)^{n_i} \bigg(1- \frac{a_i}{N}\bigg)^{TN-n_i}.
\end{equation}
Substituting $\rho(n_i)$ into $\mathcal{Q}_T(a_i,a_j,a_k)$, and using the binomial theorem to solve the sum in equation \eqref{Q integral}, we find
\begin{equation*}
    \mathcal{Q}_T(a_i,a_j,a_k) = \bigg[ \bigg(1-\frac{a_i}{N {N \choose 2}} \bigg) \bigg(1-\frac{a_j}{N {N \choose 2}} \bigg) \bigg(1-\frac{a_k}{N {N \choose 2}} \bigg)\bigg]^{TN} \simeq e^{-\frac{T}{{N \choose 2}} (a_i+a_j+a_k)}, 
\end{equation*}
where the last equivalence holds for ${N \choose 2} \gg T$.
Therefore, $\probP_T(a,a_i,a_j)$ reads as
\begin{equation}
    \probP_T(a_i, a_j, a_k) \simeq 1- e^{-\frac{T}{{N \choose 2}}(a_i +a_j + a_k)} \simeq \frac{2T}{N^2}(a_i +a_j + a_k),
\end{equation}
where we have approximated ${N \choose 2} \simeq \frac{N^2}{2}$.

Following analogous steps, the probability $\probP_T(a,a_{1}, a_{2} \ldots , a_{m})$ that $m+1$ nodes with hidden variables  $a,a_{1}, a_{2} \ldots , a_{m}$ are connected by at least one $m$-link in the aggregated HOAD network at time $T$ is
\begin{equation}
    \probP_T(a,a_{1}, a_{2} \ldots , a_{m} ) \simeq 1- e^{-\frac{T}{{N \choose m}}(a+a_{1}+a_{2}+ \ldots a_{m})} \simeq \frac{m!}{N^m}(a+a_{1}+a_{2}+ \ldots a_{m})T.
    \label{Prob approximation m}
\end{equation}
From Eq. \eqref{eq:prop-mean-approx}, the expected degree is thus
\begin{equation}
    \overline{k^{(m)}}(a)=  (a + m \langle a\rangle) T.
    \label{expected degree m}
\end{equation}

This means that, on average, the $m$-order degree of a node with activity $a$ at time $T$ is given by $Ta $ outgoing $m$-links plus $m T \langle a \rangle$ received from random neighbors.

Inserting this expression into Eq. \eqref{m-poisson-propagator}, we finally get
\begin{equation}
 g(k|a) = e^{-T(a + m\langle a\rangle )} \frac{[T(a + m\langle a\rangle )]^{k^{(m)}}}{\Gamma(k+1)}.
 \label{poisson propagator}
\end{equation}
The propagator of the order $m$ is thus functionally equivalent to the first-order one \cite{starnini2013topological}, but with a different mean. 
Again following \cite{starnini2013topological}, one can now find the explicit expression of the $m$-degree distribution of the aggregated network until time $T$, $P_T(k^{(m)})$.
\rev{Inserting Eq.\eqref{poisson propagator} into Eq. \eqref{degree distribution m order} and taking the continuum limit of latter, for $T^2 \gg k^{(m)}\gg 1$, performing a steepest descent approximation we find the asymptotic form of the degree distribution}

\begin{equation}
 P_T(k^{(m)}) \simeq \frac{1}{T}\rho (k^{(m)}/T-m\langle a\rangle ).
\end{equation}


We recall that in the approximation above we have considered a sparse hypergraph. Hence, the goodness of the approximation above depends on time $T$, which regulates the density of the higher-order network. Indeed, the HOAD model starts from disconnected nodes, which over time are connected by $m$-links, eventually percolating the network.  From Eq. \eqref{Prob approximation m}, such hypergraph sparsity condition for a HOAD model for general order $m$ is fulfilled when $T \ll \frac{N^m}{m!}$.

\subsection{Hyper-degree correlation}

We start by computing $p(a_{1}, a_{2}\ldots, a_m|a)$, the probability of randomly choosing a $m$-link made of neighbours with activities $a_{1}, a_{2}\ldots, a_m$ among all $m$-links of $a$, which is given by Eq. \eqref{eq:p-neigh}
\begin{equation}
    p(a_1, a_2 \ldots , a_m|a)= \frac{N^m \rho(a_1 )\rho(a_2 ) \ldots \rho(a_m ) }{m! \ \overline{k^{(m)}}(a)}\probP_T(a, a_{1}, a_{2}\ldots a_{m} ).
\end{equation}
By inserting the approximation for $\probP_T$ (Eq. \eqref{Prob approximation m}) and $\overline{k^{(m)}}$ (Eq. \eqref{expected degree m}), for small $T$  we find
\begin{equation}
    p(a_1, a_2 \ldots , a_m|a)  \simeq \frac{ \rho(a_1 )\rho(a_2 ) \ldots \rho(a_m ) (a + a_1 + a_2 \ldots a_m)}{(a+m\langle a\rangle)}.
    \label{prop connection}
\end{equation}

Inserting Eq. \eqref{prop connection} into Eq. \eqref{knn_h} we finally get

\begin{equation}
    \overline{k^{(m)}_{nn,T}}(a) = \frac{T}{a + m\langle a \rangle} \bigg[ \langle a^2 \rangle + (m+1)\langle a \rangle a +(m^2+m-1)\langle a \rangle^2\bigg].
    \label{knn_a final}
\end{equation}

Eq. \eqref{knn as knn_a} relates $\overline{k^{(m)}_{nn}}(k)$ with $\overline{k^{(m)}_{nn}}(a)$. Inserting Eq.\eqref{knn_a final} into Eq. \eqref{knn as knn_a}, and following equivalent steps done in \cite{starnini2013topological}, in the limit of $k^{(m)}\gg 1$ we find that the hyper-degree-degree correlation of order $m$ reads  
\begin{equation}
    \overline{k^{(m)}_{nn,T}}(k) = 1+ \frac{T^2}{k^{(m)}}\sigma^2 + (m+1 )\langle a \rangle T,
\end{equation}
\rev{being $\sigma^2 = \langle a^2 \rangle -\langle a \rangle^2 $.}
\rev{One can rewrite the latter expression in order to obtain equation (9) of the Main manuscript,}
\begin{equation}
    \frac{\overline{k^{(m)}_{nn,T}}(k) -1}{T} = (m+1 )\langle a \rangle + \frac{T}{k^{(m)}}\sigma^2 , 
\end{equation}
\rev{that gives an asymptotic form of $\overline{k^{(m)}_{nn,T}}(k)$}.

\section{Temporal percolation in hypergraphs}
In this section, we discuss temporal percolation for uncorrelated and correlated hypergraphs. We consider hypergraphs formed by hyperlinks of the same order $m$. 
Before discussing temporal percolation, we shortly introduce higher-order percolation in static hypergraphs.
We consider arbitrary hypergraphs whose nodes may be removed with probability $1-p$: when $p=0$, no node remains from the original higher-order network; by contrast, when $p=1$ all nodes are retained. 
With an approach inspired by \cite{goltsev2008percolation}, we consider the probability $x_{k}$ that, if a $m-$order hyperlink (connecting $m+1$ nodes) is attached on one side to a node with hyperdegree $k$ (where we omit the dependency in $m$ of the hyperdegree), then, following the hyperlink to its other $m$ ends, we will not end in a giant connected component.
 To this end, one needs to impose that none of the $m$ nodes (with hyperdegree $k'$) leads (through any of its remaining $k'-1$ hyperlinks) to the giant component, thus we write:
\begin{equation}
x_{k} = \bigg[ 1- p + p  \sum_{k'} P(k'| k) x_{k'}^{k'-1}\bigg]^m,
\end{equation}
 where we assume that the probability $x_{k'}$ of each of the $m$ nodes to be connected to the giant component is independent of each other, so exponentiate the same probability to the $m$.
  $P(k'| k)$ is the probability that a node with $m$-degree $k$ is connected with a node of $m$-degree $k'$.
 Close to the percolation threshold, $x_k \lessapprox  1$, hence defining $y_k = 1 - x_k \gtrapprox 0$ we write

\begin{equation}
1 - y_{k} = \bigg[ 1- p + p  \sum_{k'} P(k'| k) (1-y_{k'})^{k'-1}\bigg]^m.
\end{equation}

\rev{We then expand at the first order $(1-y_k)^{k-1}$  as $(1-y_k)^{(k-1)} \simeq 1- (k-1)y_k$, and we write}
 
\begin{equation}
    y_k = mp \sum_{k'} \textbf{B}^{(m)}_{kk'} y_{k'},
    \label{linear branching equation}
\end{equation}
where \rev{we have defined the $m$-order branching matrix as $\sum_{k'} \textbf{B}^{(m)}_{kk'} y_{k'} = \sum_{k'} P(k'| k) (k'-1)y_{k'} $ and have expanded $(1-p\sum_{k'} \textbf{B}^{(m)}_{kk'} y_{k'} )^m$}.
Moreover, following the procedure defined in \cite{starnini2014temporal}, we can easily prove that for every order $m$ the corresponding branching matrix $\textbf{B}^{(m)}_{kk'}$
 has the same element-wise form of the first order case: 
\begin{equation}
  \textbf{B}^{(m)}_{kk'} = (k'-1)\Bigg[\rho(k'-1) + \frac{\rho(k-1)}{k \rho(k)} (k' \rho(k')- \langle k \rangle \rho(k'-1)) \Bigg].
  \label{eq:branch-matr-elem-repp}
\end{equation}
Hence, we can solve Eq. \eqref{linear branching equation} as for the first-order case by considering the associated dominant eigenvalue ($\lambda_1$) equation of the branching matrix: 
\begin{equation}
    \lambda_1^2  - \langle k\rangle \lambda_1 - \langle k^2 \rangle + \langle k\rangle^2 + \langle k\rangle = 0 .
    \label{eq:lambda_thre}
\end{equation}
Notice that the last equation holds for every order and that the differences in terms of percolation time are implicitly contained in the hyper-degree momenta. 

One can also release the condition of $m$-degree correlation and by means of an analogous approach can find the percolation threshold for uncorrelated hypergraphs:

\begin{equation}
x = \bigg[ 1- p + p  \sum_{k'} \frac{k' \rho(k')}{\langle k \rangle} x^{k'-1}\bigg]^m,
\end{equation}
where the probability $x_k =x$ does not depend on $k$ anymore. Defining $x = 1-y$ and developing till the first order we find the $m$-order Molloy-Reed criterion:
\begin{equation}
y =  m \ p  \frac{\langle k^2 \rangle- \langle k \rangle}{\langle k \rangle}  \ y,
\end{equation}

and for $p = 1$, last expression has non-trivial solution in x for:
\begin{equation}
    \frac{\langle k^2 \rangle- \langle k \rangle}{\langle k \rangle} > \frac{1}{m}
\end{equation}
In the uncorrelated case, we found that the results were consistent with those by Sun et al \cite{sun2021higher}.

\subsection{Temporal percolation in the HOAD model}
We now consider the related problem of temporal percolation in the HOAD model. To this end, we utilized calculations equivalent to those presented by Starnini et al. \cite{starnini2014temporal}, but we report them here for clarity.
To find the percolation time we need to express the hyperdegree momenta as a function of the activity variable momenta. 
We can write the hyperdegree momenta $\langle k^n\rangle_T$ at a time $T$ with respect to the time-dependent propagator $g_T(k|a)$ as

\begin{equation}
\langle k^n\rangle_T = \sum_a \rho(a) \sum_k k^n g_T(k|a).
\label{eq:momenta-relation-1}
\end{equation}
Since the propagator has the form of a Poisson distribution, the momenta of the degree distribution simply read as

\begin{equation}
    \langle k^n\rangle_T = \sum_{i=1}^n {n\brace i} T^i \kappa_i,
    \label{eq:momenta-relation-2}
\end{equation}

where ${n\brace i}$ are the Stirling numbers of the second kind and 
\begin{equation}
    \kappa_i = \sum_a \rho(a) (a+ m\langle a \rangle)^i= \sum_{j=0}^i \binom{i}{j} \langle a^j \rangle (m\langle a  \rangle )^{i-j}.
    \label{eq:momenta-relation-3}
\end{equation}

Explicitly, $\langle k \rangle_T$ and $\langle k^2 \rangle_T$ can be written as a function of the first two activity momenta as 
\begin{equation}
    \langle k \rangle_T = T \kappa_1 = T(m+1) \langle a \rangle, 
\end{equation}

\begin{equation}
    \langle k^2 \rangle_T = T \kappa_1 +T^2 \kappa_2 = T(m+1) \langle a \rangle + [\langle a^2 \rangle +  (m^2 +2m) \langle a \rangle^2] T^2 .
\end{equation} 
Using the last expressions we can find an analytical approximation for the percolation time of hypergraphs with no hyperedgree correlations as a function of the different orders of interactions present in the hypergraphs.
For instance, we can solve Eq. \eqref{eq:lambda_thre} by means of Eq. \eqref{eq:momenta-relation-1} \eqref{eq:momenta-relation-2} \eqref{eq:momenta-relation-3}, leading to the following formula for the percolation time of uncorrelated hypergraphs for any order $m$:

\begin{equation}
    T^{(m)}_c = \frac{2}{(m+1)\langle a \rangle  + \sqrt{(m^2+2m-3)\langle a \rangle^2 +4\langle a^2 \rangle} }.
    \label{correlated percolation last equation}
\end{equation}

Notice that the last equation holds for every order and that the differences in terms of percolation time are implicitly contained in the hyper-degree momenta. 
Equivalently, for uncorrelated hypergraphs, exploiting again the relation between the momenta $\langle (k^{(m)} )^n\rangle$ and $\langle a^n\rangle$ (Eq. \eqref{eq:momenta-relation-1},\eqref{eq:momenta-relation-2},\eqref{eq:momenta-relation-3}), we find a general prediction of the percolation time for uncorrelated temporal higher-order networks: 

\begin{equation}
    T^{(m)}_{unc}=\frac{\left(m + 1\right)\langle a \rangle }{m(m +2) \langle a \rangle^{2}  + \langle a^2 \rangle}.
\end{equation}

\begin{figure}[tbp]
    \centering
    \includegraphics[width =.9\columnwidth]{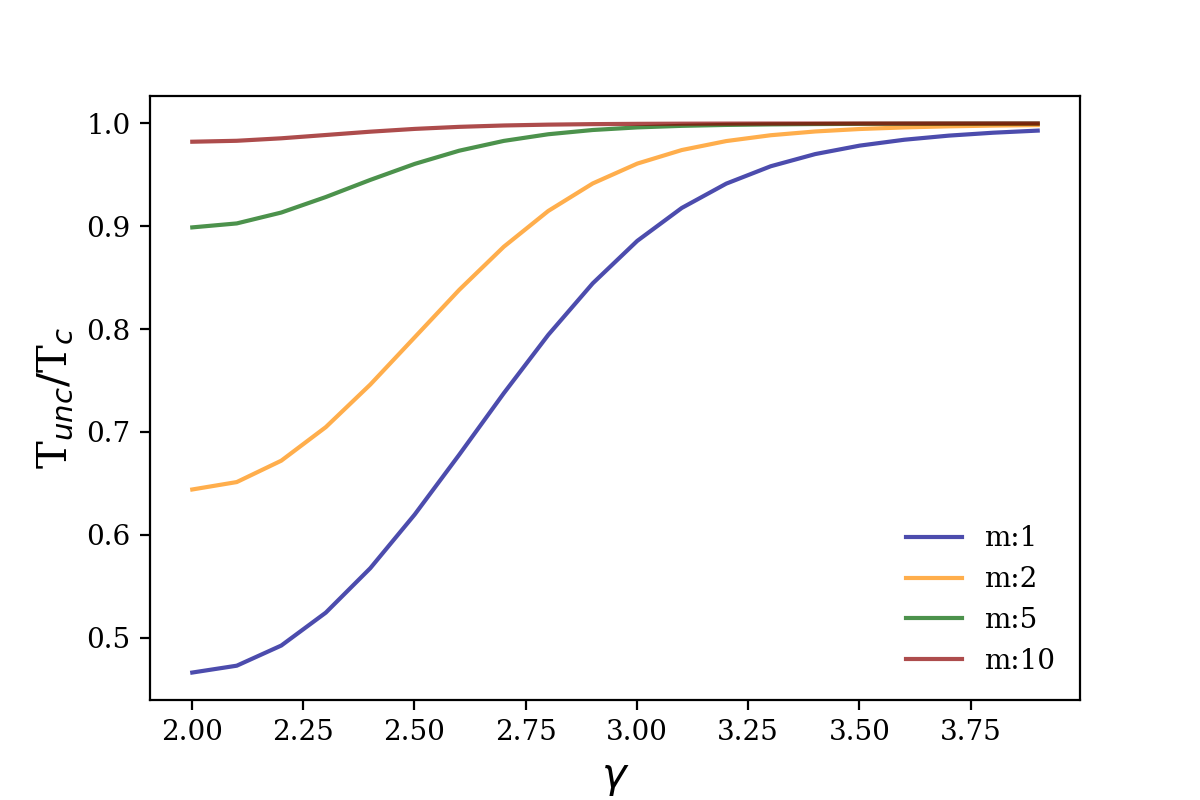}
    \caption{\small \textbf{$\frac{T_{unc}}{T_c}$ as a function of $\gamma$.}
    Network size $N = 10^6$, orders $m=[1,2,5,10]$ and $\gamma \in [2,4]$. 
    The activity distributions $\rho(a)$ are power-law distributions with $\epsilon = 10^{-3}$. }
    \label{T_ratio_gamma}
\end{figure}

\begin{figure}[tbp]
    \centering
    \includegraphics[width =.9\columnwidth]{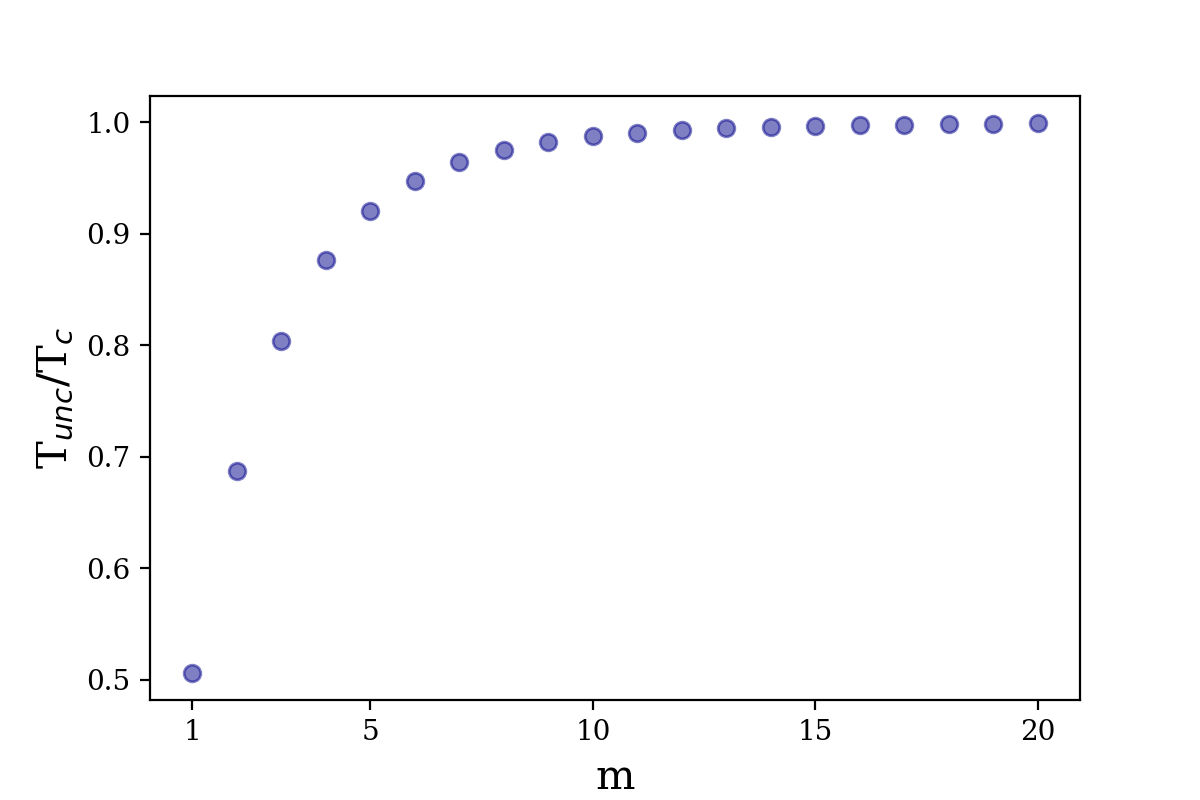}
    \caption{\small \textbf{$\frac{T_{unc}}{T_c}$ as a function of $m$.}
    Network size $N = 10^6$, orders $m=[1,\ldots,20]$. 
    The activity distributions $\rho(a)$ have the same power-law form for every order with exponent $\gamma = 2.25$, with $\epsilon = 10^{-3}$. }
    \label{T_ratio_m}
\end{figure}

We now compare the analytical prediction of percolation time for uncorrelated and correlated hypergraphs by plotting the ratio $\frac{T^{(m)}_{unc}}{T^{(m)}_c}$ (Eq.\eqref{eq:T_ratio}) for different orders $m$ and different power-law exponent $\gamma$ of the activity distribution.
\begin{equation}
    \frac{T^{(m)}_{unc}}{T^{(m)}_c}=\frac{\left(m + 1\right)\langle a \rangle\left(\left(m + 1\right)\langle a \rangle + \sqrt{(m^2+2m-3)\langle a \rangle^2 +4\langle a^2 \rangle}\right)}{2 \left( m(m +2) \langle a \rangle^{2}  + \langle a^2 \rangle \right)}
    \label{eq:T_ratio}
\end{equation}
We observe that the uncorrelated case underestimates the percolation time with respect to the correlated one for every $m$ and $\gamma$, Fig. \ref{T_ratio_gamma} and \ref{T_ratio_m}. 
The difference between $T^{(m)}_{unc} $ and $T^{(m)}_{c}$ is maximum for strongly heterogeneous hypergraphs (Fig. \ref{T_ratio_gamma}), for instance $\gamma = 2$.
For large $m$, $m \ll 1$, one can see that $T^{(m)}_{unc} \rightarrow T^{(m)}_{c}$, see   Fig. \ref{T_ratio_m}.

We test the validity of the 
theoretical prediction of the percolation threshold for correlated HOAD networks by running extensive numerical simulations.
Figure \rev{2 (a) of the main manuscript and Figure \ref{Percolation SM}} show the growth of the giant component size $S$ over time and the peak of its variance, $\sigma(S)^2$,  indicating the estimated percolation threshold, for several orders $m$. 
The theoretical prediction of the percolation threshold is shown to have a decent agreement with numerical results, yet they do not exactly coincide.

 \newpage
 \clearpage
 \newpage

\section{Supplementary Figures}

Supplementary Figures supporting theoretical predictions for the hyper-degree distribution, hyper-degree correlations, and percolation thresholds for different orders $m$.



\begin{figure}[h!]
    \centering
    \includegraphics[width =.9\columnwidth]{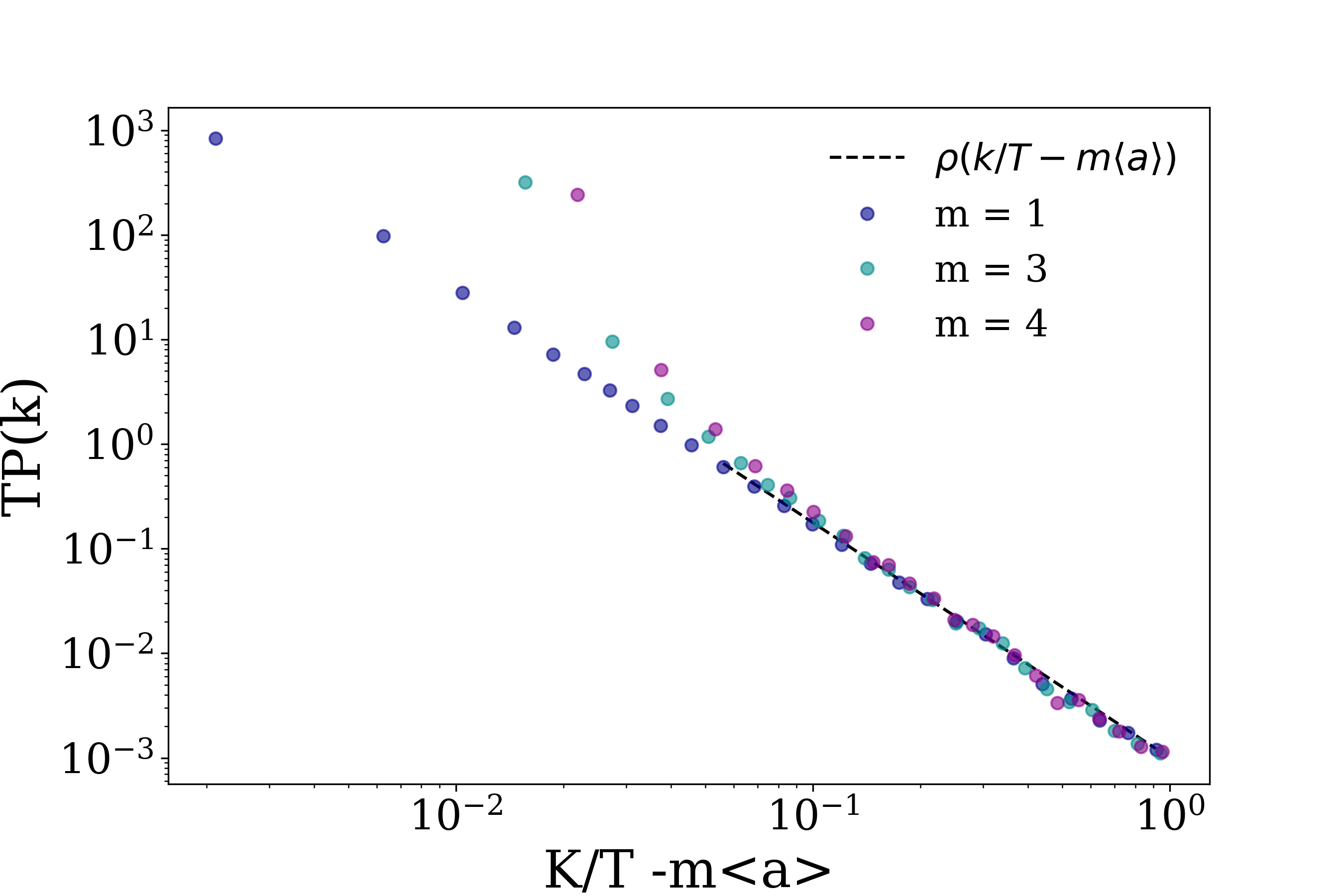}
    \caption{\small \textbf{Hyper-degree distribution $P_T(k^{(m)})$ of HOAD networks.}
    Network size $N = 10^6$, orders $m=1,3,4$, integration time $T=10^3$. 
    The activity distributions $\rho(a)$ have the same power-law form for every order with exponent $\gamma = 2.25$, with $\epsilon = 10^{-3}$. }

\end{figure}

\begin{figure}[h!]
    \centering
    \includegraphics[width =.9\columnwidth]{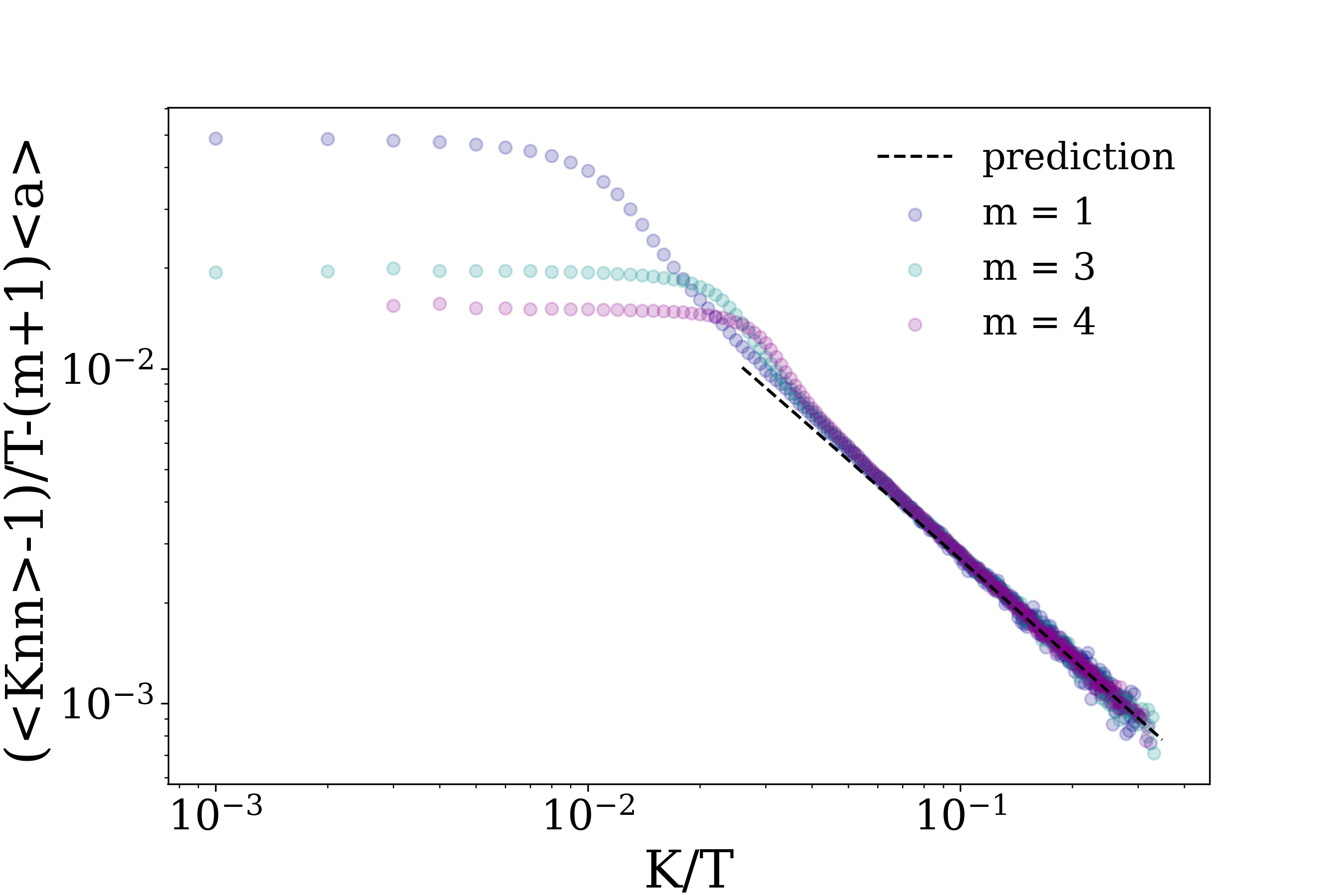}
    \caption{\small \textbf{Hyper-degree correlations $\overline{k^{(m)}_{T,nn}}(k)$ of HOAD networks.}
    Network size $N = 10^6$, orders $m=1,3,4$, integration time $T=10^3$. 
    The activity distributions $\rho(a)$ have the same power-law form for every order with exponent $\gamma = 2.25$, with $\epsilon = 10^{-3}$.}
\end{figure}

\begin{figure}[h!]
    \centering
    \includegraphics[width =.9\columnwidth]{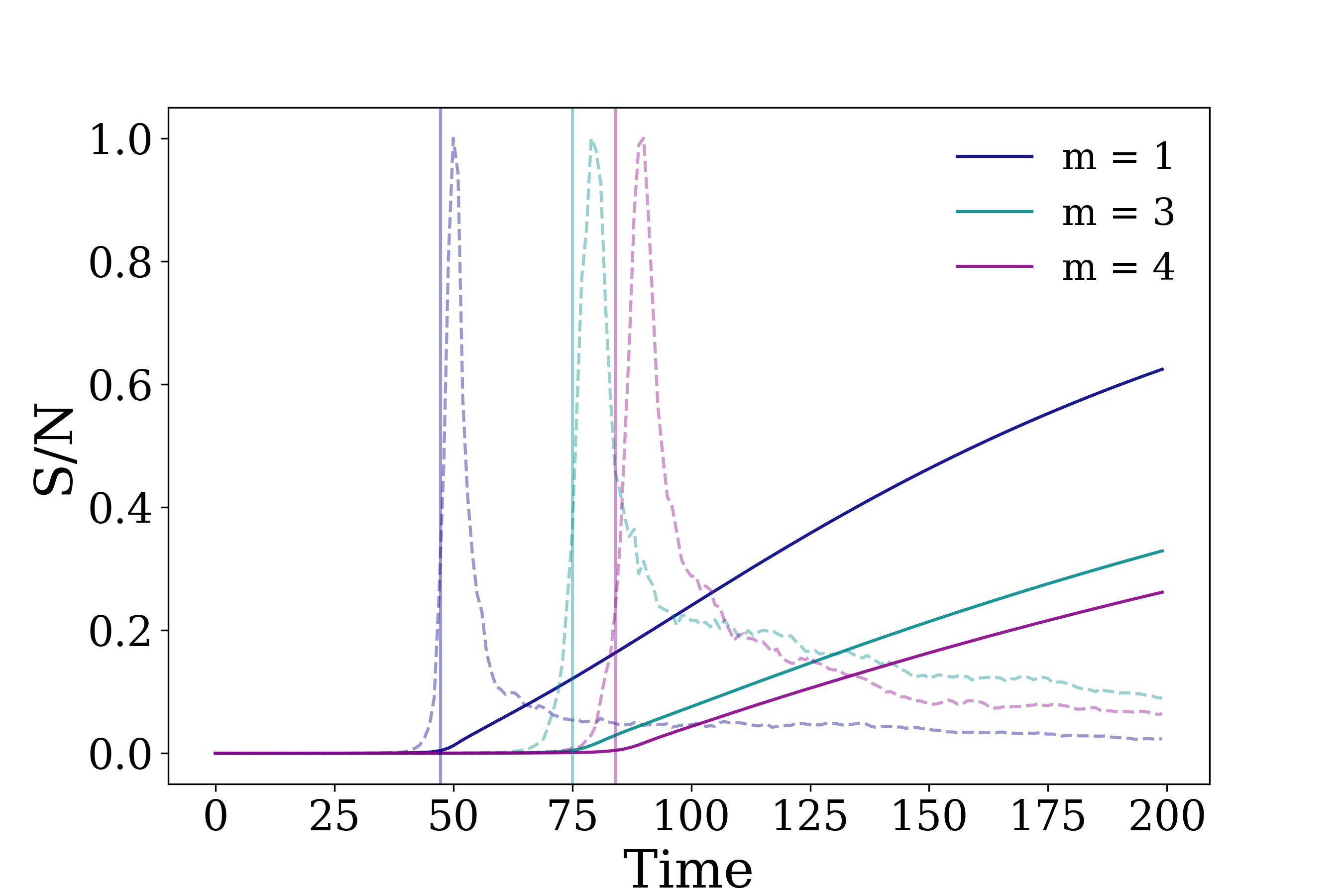}
\caption{\small \textbf{Percolation time.} Orders $m = 1, 3, 4$. 
    Giant component size $S/N$ (continuous line) and the peak of its variance $\sigma(S)^2$ (dashed line) over time. 
    Results are averaged over $10^2$ runs.}
    \label{Percolation SM}
\end{figure}

\clearpage
\newpage

\section{Empirical Data}
\subsection{Data Description}
In this study, we utilized two data sets, namely \textit{coauth-mag-geology} and \textit{coauth-mag-history}, obtained from the repository available at \url{https://gitlab.com/complexgroupinteractions/xgi-data}. These datasets consist of timestamped higher-order interactions, where each interaction is a set of nodes. Specifically, the \textit{coauth-mag-geology} dataset comprises publications tagged with Geology in the Microsoft Academic Graph, while the \textit{coauth-mag-history} dataset consists of publications tagged with History in the same dataset. 
Nodes within these data sets correspond to authors, and the timestamps indicate the publication year. The projected graphs are weighted undirected networks, reflecting the co-occurrence of author pairs within higher-order interactions. 
For the sake of our analysis, we focused on interactions with a maximum of 11 nodes ($10^{th}$ order).

\subsection{Quantifying higher-order and first-order activity distributions}

The higher-order activity potential of individuals has been extracted from data as detailed in the definition of the HOAD model. 
Specifically, we counted the number of interactions each node participated in for different orders and divided by the total number of interactions across all orders. 
The first-order activity potential of individuals has been extracted from data as detailed in the definition of the activity-driven model \cite{perra2012activity}.

We then directly compare the higher-order percolation threshold with the first-order one.
To this aim, we project all interactions into
the first order, thus representing higher-order data as
a simple network, and measure the activity potential in this case.  
We note that, in order to meaningfully compare the two cases, 
the first-order activities of nodes must be multiplied by the factor ${m+1 \choose 2}$, indicating the number of equivalent links included in a $m-$order interaction. 
In this way, we ensure that, at any given time $T$, the simple activity-driven network and the higher-order activity-driven network projected to the first-order have the same number of links.

\end{document}